\documentclass[12pt]{article}

\begin{document}

\newfont{\elevenmib}{cmmib10 scaled\magstep1}

\newcommand{\YUKAWAmark}{\elevenmib
            Yukawa\hskip1mm Institute\hskip1mm Kyoto}

\newcommand{\preprint}{	\begin{flushleft}						 
			     \YUKAWAmark	
                        \end{flushleft}	  \vspace{-0.95cm}	
		        \begin{flushright} \normalsize  \sf		
		      	     YITP-96-25\\ quant-ph./9607012	\\
                             June 1996
                        \end{flushright}}

\baselineskip=20pt

\renewcommand{\theequation}{\arabic{section}.\arabic{equation}}

\title{  \preprint  \Large \bf \ \\ \ \\ 
         Generalized Binomial States: Ladder Operator Approach \\ \ 
\\ 
         }

\author{ Hong-Chen Fu\thanks{   On leave of absence from Institute
                                of Theoretical Physics, Northeast 
                                Normal University, Changchun 130024, 
                                P.R.China.
                                E-mail: hcfu@yukawa.kyoto-u.ac.jp }\ \
         and Ryu Sasaki \\ \ \\
         {\normalsize \it       Yukawa Institute for Theoretical 
                                Physics, Kyoto University,}\\
         {\normalsize \it       Kyoto 606-01, Japan}}

\date{\ }

\maketitle

\vspace{0.3cm}

\begin{abstract}
We show that the binomial states (BS) of Stoler {\it et al.}
admit the ladder and displacement operator formalism. By
generalizing the ladder operator formalism we propose an
eigenvalue equation which possesses the number and the squeezed
states as its limiting solutions. The explicit forms of the solutions,
to be referred to as the {\it generalized binomial
states} (GBS), are given. Corresponding
to the wide range of the eigenvalue spectrum these GBS have as
widely different properties. Their
limits to number and {\it squeezed} states are investigated in
detail. The time evolution of BS is obtained as a special
case of the approach.
\end{abstract}

\vspace{1.6cm}

\begin{center} \large \sf Journal of Physics A:
Mathematical and General\\ \vspace{0.1cm}\tiny (Accepted for publication) 
\end{center}

\newpage

\section{Introduction}
\setcounter{equation}{0}

The number and the coherent states of quantized radiation field
play important roles in quantum optics and are extensively studied
\cite{noch}. The binomial states (BS) introduced by Stoler,
Saleh and Teich in
1985 \cite{stol}, interpolate between the {\it most nonclassical}
number states and the {\it most classical} coherent states, and
reduce to them in two different limits. Some of their properties
\cite{stol,leee,barr}, methods of generation \cite{stol,leee,datt},
as well as their interaction with atoms \cite{josh}, have been
investigated in the literature. The BS is defined as a linear
superposition of number states in an $M$-dimensional subspace
\begin{equation}
  |\eta,\,M\rangle=\sum_{n=0}^{M}\beta_n^M(\eta)|n\rangle,
  \label{bs}
\end{equation}
where $\eta$ is a real parameter satisfying $0<\eta<1$
(``probability''), and
\begin{equation}
  \beta_n^M(\eta)=\left[\left(\begin{array}{c}M\\n\end{array}
  \right) \eta^n (1-\eta)^{M-n}\right]^{1/2}.
  \label{bsdist}
\end{equation}
The name `binomial state' comes from the fact that their photon 
distribution $|\langle n|\eta,\,M\rangle|^2=|\beta_n^M(\eta)|^2$
is simply a binomial distribution with probability $\eta$.
In the two limits $\eta \to 1$ and  $\eta \to 0$ (in both cases 
``definite probability'') it reduces to number states:
$|1,\,M\rangle=|M\rangle$ and $|0,\,M\rangle=|0\rangle$,
respectively. In a different limit of $M\to \infty, \  \eta\to 0$
with $\eta M =\alpha^2$ fixed ($\alpha$ real constant),
$|\eta,\,M\rangle$ reduces to the  coherent states (not the most
general ones, only those with real amplitude $\alpha$), which
corresponds to the Poisson distribution in probability theory
\cite{Feller}. It is well known that the binomial distribution
tends to the Poisson distribution in the above limit\cite{Feller}.
The notion of BS was also generalized to the intermediate
number-squeezed states \cite{bas1} and the number-phase states
\cite{bas2}, as well as their $q$-deformation \cite{fann}.

It is well known that the number and the coherent states are
the eigenstates of the number operator $N$ and the annihilation 
operator $a$, respectively. So we naturally ask if BS is an
eigenstate of a proper linear combination of the number operator
and the (density-dependent) annihilation operator, or in other
words, if it admits a {\it ladder operator definition}. The answer
is positive. In Sec.2 we show that BS is the eigenstate of the
combination of number operator $N$ and raising operator
$J^+_M=\sqrt{M-N}\,a\, $ of SU(2) via its Holstein-Primakoff 
realization.
This ladder operator formalism enables us to easily derive their
{\it displacement operator} formalism. The result shows that BS
is in fact a {\it special} SU(2) coherent states, as noted in
\cite{fann}.

In Sec.\,3, we generalize the ladder operator approach of
BS towards the generalized binomial state (GBS) in the sense
that they reduce to the number and coherent, squeezed states
in certain limits. Recall that the squeezed states of the
radiation field are the eigenstates of a linear combination
of its creation and annihilation operators. So we replace the
linear combination of raising and lowering operators of SU(2)
instead of the raising operator in the ladder operator form
of binomial states and thus obtain an eigenvalue equation of
proper linear combination of {\it all} generators of SU(2)
(to be referred to as GBS equation for convenience). The GBS
equation is exactly solved using a method developed in the
investigation of the squeezed states of SU(1,1) algebras
\cite{fus3} and its $M+1$ distinct eigenvalues and
corresponding eigenstates are found. In Sec.\,4 we show that
these solutions degenerate to the number, coherent and squeezed
states in different limits. In Sec.\,5 we point out that BS and
its time evolution are the special case of GBS equation
with a special eigenvalue. We conclude in Sec.\,6.

\section{Ladder operator approach to BS}
\setcounter{equation}{0}

Let us first consider the ladder operator formalism of BS.
To this end we suppose that the BS is an eigenstate of a linear
combination of the number operator $N$ and a density-dependent
annihilation operator $f(N)a$, namely,
\begin{equation}
    \left[\mu N + \nu f(N) a \right]|\eta,\,M\rangle =
    \rho |\eta,\,M\rangle, \label{an}
\end{equation}
where constants $\mu$, $\nu$ $\rho$ and a function $f(N)$ are
to be determined. Taking into account that $\mu=1$ in the limit
$\eta \to 1$ and $\mu=0$ in the coherent state limit
$\eta\to 0$, we can simply choose $\mu=\sqrt{\eta}$. Then,
inserting the explicit BS (\ref{bs}) into (\ref{an}),
we obtain the following equations
\begin{eqnarray}
    &&\sqrt{\eta^M} (\rho-\sqrt{\eta} M)=0, \nonumber \\
    &&\sqrt{1-\eta}(\rho-n\sqrt{\eta})=
      \nu \sqrt{\eta(M-n)}f(n),\ \ \
      (n=0,1,\cdots,M-1),
\end{eqnarray}
from which we find
\begin{equation}
    \rho=\sqrt{\eta} M, \ \ \ \nu=\sqrt{1-\eta}, \ \  \
    f(N)=\sqrt{M-N}.
\end{equation}
Substituting these results into \,(\ref{an}), we obtain the
ladder operator formalism of BS
\begin{equation}
   \left[\sqrt{\eta} N + \sqrt{1-\eta}\sqrt{M-N}\,a \right]
   |\eta,\,M\rangle
   =\sqrt{\eta}M|\eta,\,M\rangle.  \label{eig}
\end{equation}
It is interesting that the operators appearing in the above
equation (\ref{eig}) are the well-known Holstein-Primakoff
realization of Lie algebra SU(2):
\begin{equation}
    J_M^0 \equiv \frac{M}{2}-N,\ \ \
    J_M^+ \equiv \sqrt{M-N}\,a,\ \ \
    J_M^- \equiv a^{\dagger}\sqrt{M-N}. \label{hp}
\end{equation}
In terms of the SU(2) generators, \,(\ref{eig}) is rewritten as
\begin{equation}
    \left[\sqrt{\eta}J_M^0-\sqrt{1-\eta}\,J_M^+\right]|\eta,\,M\rangle
    =-\frac{\sqrt{\eta}M}{2}|\eta,\,M\rangle. \label{27}
\end{equation}
This characterization of the BS in terms of the SU(2)
operators is consistent with the original definition (\ref{bs}), 
(\ref{bsdist}) in the two limits of
``definite probability'' $\eta \to 1$ and  $\eta \to 0$: 
\begin{displaymath}
     N |1,\,M\rangle=M|1,\,M\rangle,\qquad a|0,\,M\rangle=0,
\end{displaymath}
respectively. To achieve the coherent state in some limit, we 
multiply $\sqrt{\eta}$ on both sides of (\ref{eig}). Then,
considering the limit $M\to\infty$ and $\eta\to 0$ with  fixed
$M\eta=\alpha^2$  ($\alpha$ is a real constant) for finite $n$,
we arrive at the equation
\begin{equation}
    a|0,\infty\rangle =\alpha\, |0,\infty\rangle,
\end{equation}
which is nothing but the ladder (annihilation) operator definition
of coherent state.

Here we would like to remark that the binomial state is only one
eigenstate of the operator $\sqrt{\eta}N+\sqrt{1-\eta}J_M^+$
corresponding to the eigenvalue $\sqrt{\eta}M$. This
operator generally has $M+1$ eigenvalues and eigenstates since it
is in fact an $(M+1)\times(M+1)$ matrix. In Sec.\,5, the complete
eigenvalues and eigenstates will be presented.

From the ladder operator form of BS we can easily derive its
displacement operator form. For this purpose, we identity
$\sqrt{\eta}=\sin  r$, $\sqrt{1-\eta}=\cos r$, $0<r<\pi/2$.
Then (\ref{27}) can be rewritten as
\begin{equation}
   (\cos r J^+_M - \sin r J^0_M)|\eta,M\rangle=\frac{1}{2}
   M\sin r |\eta,M\rangle. \label{todis}
\end{equation}
Comparing (\ref{todis}) with the atomic coherent states and
its ladder operator 
form\footnote{Symbols here are different from those in
          \cite{atom}: $J^{\pm}_M\to J_{\mp},\ J^0_M\to
          -J_{z}$, $M\to 2J$ and $|0\rangle \to
          |-J\rangle$.}
in \cite{atom}, we find that the BS can be written as
\begin{equation}
   |\eta,M\rangle=e^{-r(J^+_M-J^-_M)}|0\rangle.
   \label{laddform}
\end{equation}
So BS can be viewed as a special SU(2) coherent state, as noted
in \cite{fann}.

\section{GBS equation and exact solutions}
\setcounter{equation}{0}

On the basis of the above analysis, we shall propose a more
general eigenvalue equation, the GBS equation, which possesses
the number and the squeezed states as its limiting solutions,
and present its exact solutions in this section.

From the discussions in Sec.2 we see that in the limit $\eta\to 0$
and $M\to \infty$ with fixed $\eta M =\alpha^2$, 
$\sqrt\eta J^+_M \to \alpha a$. In fact,
we also have $\sqrt\eta J^-_M\to \alpha a^{\dagger}$ in the same 
limit.
 Recall that
the squeezed states of a single mode radiation field can be defined
as the eigenstates of the operator $\mu a +\nu a^{\dagger}$,
where two complex numbers $\mu$ and $\nu$ satisfy $|\nu/\mu|
<1$. So, to achieve the GBS, we should replace $J^-_M$ in
the (\ref{eig}) by the operator
$\mu J^+_M +\nu J^-_M$.
 (Note that $|\nu/\mu|<1$ was necessary for the 
convergence of an infinite series for the squeezed state. 
In the GBS case such a constraint is not necessary since 
the operators are finite matrices.) 
In summary, we propose the following equation
\begin{equation}
   \left[\sqrt{1-\eta}(\mu J^+_M +\nu J^-_M) -\sqrt{\eta}J^0_M\right]
   |\beta,\delta\rangle=\delta|\beta,\delta\rangle, \label{three}
\end{equation}
where $\beta=\{\mu,\nu,\eta,M \}$ and $\mu\neq 0$
without loss of generality.

Equation (\ref{three}) is an eigenvalue equation for an $(M+1)
\times(M+1)$-matrix. So, generally speaking, it has up to $M+1$ 
 different eigenvalues. If we expand the state
$|\beta\rangle$ in terms of the number states, (\ref{three})
will lead to a recursion relation with three terms, which is
difficult to solve in its full generality. Here we shall use a
method used in the investigation of the squeezed states of su(1,1)
algebra \cite{fus3}. We write the state $|\beta\rangle$ in the
form
\begin{equation}
   |\beta,\delta\rangle = D(\zeta)\|\beta,\delta\rangle,\ \ \
   D(\zeta) = \exp(\zeta J^+_M -\zeta^* J^-_M),
\end{equation}
where the parameter 
\begin{equation}
   \zeta = r e^{i\theta}  \label{zetadef}
\end{equation} 
will be determined later. Then by making use of the following 
relations
\begin{eqnarray}
&&   D^{-1}(\zeta)J^+_M D(\zeta)=J^+_M \cos^2 r -J^-_M\sin^2 r
     e^{-2i\theta}-J^0_M \sin(2r)e^{-i\theta},  \nonumber \\
&&   D^{-1}(\zeta)J^-_M D(\zeta)=J^-_M \cos^2 r -J^+_M\sin^2 r
     e^{2i\theta}-J^0_M \sin(2r)e^{i\theta},      \nonumber \\
&&   D^{-1}(\zeta)J^0_M D(\zeta)=\frac{1}{2}( 
     J^+_M e^{i\theta}+J^-_M e^{-i\theta})\sin(2r)+J^0_M\cos(2r),
\end{eqnarray}
we obtain an equation for $\|\beta,\delta\rangle$
\begin{equation}
    (A_+ J^+_M + A_- J^-_M - A_0 J^0_M)\|\beta,\delta\rangle =
    \delta \|\beta,\delta\rangle, \label{thr2}
\end{equation}
where
\begin{eqnarray}
&&    A_+ = \sqrt{1-\eta}(\mu \cos^2 r -\nu \sin^2 r e^{2i\theta})
            -\frac{1}{2}\sqrt{\eta} e^{i\theta}\sin(2r), \nonumber \\
&&    A_- = \sqrt{1-\eta}(\nu \cos^2 r -\mu \sin^2 r e^{-2i\theta})
            -\frac{1}{2}\sqrt{\eta} e^{-i\theta}\sin(2r), \nonumber \\
&&    A_0 = \sqrt{1-\eta}(\mu e^{-i\theta} +\nu e^{i\theta})\sin(2r)
            +\sqrt{\eta}\cos(2r). \label{aaa}
\end{eqnarray}
As an important step, we choose $\zeta$ such that $A_-=0$, namely,
\begin{equation}
     \cos^2 r \left[\sqrt{1-\eta}(\nu-\mu \tan^2 r e^{-2i\theta})-
     \sqrt{\eta}\tan r e^{-i\theta}\right]=0,
\end{equation}
where we have used the fact $\cos r \neq 0$ (otherwise,
$A_-=-\sqrt{1-\eta}\mu e^{2i\theta}\neq 0$). So $\zeta$ is
determined by
\begin{equation}
    \mu\sqrt{1-\eta} \Delta^2 + \sqrt{\eta}\Delta -\sqrt{1-\eta}
    \nu = 0, \ \ \ \  \Delta=e^{-i\theta}\tan r, \label{cons1}
\end{equation}
and in this case  (\ref{thr2}) is reduced to
\begin{equation}
    A_+ J^+_M \|\beta,\delta\rangle = (\delta+ A_0 J^0_M)
    \|\beta,\delta\rangle.    \label{twoterm}
\end{equation}
Let $\| \beta,\delta\rangle=\sum_{n=0}^{M}C_n |n\rangle$ and
insert it into (\ref{twoterm}). We obtain
\begin{eqnarray}
    && C_M(\delta-MA_0/2)=0, \label{cm1}\\
    && C_{n+1}\sqrt{(n+1)(M-n)}A_+ = C_n (\delta+A_0M/2-A_0n),
       \nonumber \\
    && \hspace{2cm}   (n=0,1,\cdots,M-1).\label{cm2}
\end{eqnarray}
From (\ref{cm1}) we have two possibilities: $\delta-MA_0/2=0$
or $C_M=0$. In the first case, we can determine one eigenvalue
$\delta_M=MA_0/2$ and its corresponding eigenstate from
(\ref{cm2}). If $C_M=0$, we still have two possibilities:
$\delta+A_0M/2-A_0(M-1)=0$ or $C_{M-1}=0$. Performing the analysis
in the same way, we obtain all eigenvalues and eigenstates.
In the general case, $C_M=\cdots=C_{k+1}=0$ and $C_{k}\neq 0$,
$k=0,1,\cdots,M$, (\ref{cm1}) and (\ref{cm2}) reduce to
\begin{eqnarray}
   && C_{k}(\delta+MA_0/2-A_0 k)=0, \label{cm3}\\
   && C_{n+1}\sqrt{(n+1)(M-n)}A_+ = C_n (\delta+A_0M/2-A_0n),
      \nonumber \\
   && \hspace{2cm}   (n=0,1,\cdots,k-1).\label{cm4}
\end{eqnarray}
From (\ref{cm3}) we obtain the eigenvalues
\begin{equation}
   \delta_k = \frac{1}{2}A_0 (2k-M),\ \ \ \
   k=0,1,\cdots,M,   \label{delk}
\end{equation}
which are non-degenerate. Substituting (\ref{delk}) into
(\ref{cm4}), we can determine the corresponding eigenstates.
We evaluate them for two typical cases:

{\it Case 1}. If $A_+$ is non-vanishing, the corresponding eigenstates
are obtained as
\begin{equation}
   \|\beta,\delta_k\rangle=C_0\sum_{n=0}^{k}\left(
   \begin{array}{c}k\\n\end{array}\right)
   \left(\begin{array}{c}M\\n\end{array}
   \right)^{-1/2} A_0^n A_+^{-n}|n\rangle.    \label{eista1}
\end{equation}
Those eigenstates of different eigenvalues are linearly independent.
Using the {\it identity} method developed in \cite{fus2}, we can
rewrite the eigenstates $\|\beta,\delta_k\rangle$ in the {\it
exponential form}
\begin{equation}
   \|\beta,\delta_k\rangle=C_0'\exp\left\{\frac{A_0}{A_+}\sqrt{
   \frac{k-N+1}{M-N+1}}\ J_k^-\right\}|0\rangle, \label{eista2}
\end{equation}
where we have used the following identity
\begin{equation}
   (f(N)a^{\dagger})^n |0\rangle=(a^{\dagger})^n f(1)f(2)\cdots f(n)
   |0\rangle  \ \ \ \ \mbox{with} \ \ \ \
   f(N)=\frac{k-N+1}{\sqrt{M-N+1}}.
\end{equation}
Let us discuss the cases $k=0$ and $M$ in more detail. In the case
of $k=M$, (\ref{eista1}) becomes
\begin{equation}
   \|\beta,\delta_M\rangle=\sum_{n=0}^{M}\left[\left(
   \begin{array}{c}M\\n \end{array}\right)
   \eta'^n (1-\eta')^{M-n}\right]^{1/2}
   \ e^{in(\theta_0-\theta_+)}|n\rangle,   \label{eista3}
\end{equation}
where
\begin{equation}
   \eta'=\frac{|A_0|^2}{|A_0|^2+|A_+|^2},\ \ \ \
   A_0=|A_0|e^{i\theta_0},\ \ \ \
   A_+=|A_+|e^{i\theta_+}.
\end{equation}
It is obvious that the state (\ref{eista3}) is a binomial state
with a very special phase structure.
So the eigenstate $|\beta,\delta_M\rangle$ is finally obtained as
\begin{equation}
   |\beta,\delta_M\rangle=D(\zeta)\|\beta,\delta_M\rangle,
\end{equation}
which is the {\it displaced binomial state}. In the case of
$k=0$, it is easy to see from (\ref{eista1}) that
$\|\beta,\delta_0\rangle =|0\rangle$ and
\begin{equation}
   |\beta,\delta_0\rangle=D(\zeta)|0\rangle
\end{equation}
is a binomial state with a phase (see (\ref{laddform})).

{\it Case 2}. Next let us consider a very special case
  $A_+=0$. It is obvious that the spectrum is given 
by (\ref{delk}) but the eigenstates are determined by
\begin{equation}
   C_n (\delta_k+A_0M/2-A_0n)=0, \ \ \  
   (n=0,1,\cdots,k-1), \ \ \ C_k\neq 0,   \label{eista4}
\end{equation}
from which we find that $C_n=0 \ (n\neq k)$, namely, 
$\|\beta,\delta_k\rangle=|k\rangle$, the number states. Therefore,
we finally obtain
\begin{equation}
   |\beta,\delta_k\rangle=D(\zeta)|k\rangle,
\end{equation}
where $\zeta$ should satisfy (\ref{cons1}) and an additional
equation ($A_+=0$)
\begin{equation}
    \sqrt{1-\eta}\nu \Delta^{*2}+
    \sqrt{\eta}\Delta^* -\sqrt{1-\eta}\mu=0,\quad 
    \Delta^*=e^{i\theta}\tan r.
    \label{cons2}
\end{equation}
By comparing (\ref{cons1}) and (\ref{cons2}) we find these
two equations are simultaneously satisfied if $\eta=1$ or
\begin{displaymath}
	\mu=\nu^*
\end{displaymath}
In next section we shall consider the limit $\eta\to 1$ in which
the binomial states tend to the number states.

In conclusion, we have found that (\ref{three}) has $M+1$ distinct
eigenvalues and corresponding linearly independent eigenstates
and GBS equation (\ref{three}) finally takes the form
\begin{equation}
   \left[\sqrt{1-\eta}(\mu J^+_M +\nu J^-_M) -
   \sqrt{\eta}J^0_M\right]|\beta,\delta_k\rangle=
   \frac{A_0}{2}(2k-M)|\beta,\delta_k\rangle.\label{ness}
\end{equation}

\section{Limit to number and squeezed states}
\setcounter{equation}{0}
In this section we discuss the limiting cases of the GBS obtained 
in the previous section.
Let us first consider the limit $\eta\to 1$. In this case 
(\ref{cons1}) requires  $\Delta=0$ or $\sin r=0$ 
($r=m\pi$: $m$ integers) and $A_0\to1$ and (\ref{ness}) reduces to
\begin{equation}
   N|\mu,\nu,1,M;\delta_k\rangle=
   k|\mu,\nu,1,M;\delta_k\rangle.
   \label{limnum}
\end{equation}
Namely, $|\beta,\delta_k\rangle$ goes to the number state
$|k\rangle$.

The same conclusion can be reached by a different path.
From the disentangling theorem \cite{trua}
\begin{equation}
   D(\xi)= \exp\left(\xi J_M^+ -\xi^* J_M^-\right)=
    \exp(-\tau^* J_M^-)\exp[-\ln(1+|\tau|^2) J_M^0 ]    
    \exp(\tau J_M^+), \label{disen}
\end{equation}
where $\xi=|\xi|e^{-i\phi}, \  \tau=e^{-i\phi}\tan|\xi|$,
we have $D(m\pi e^{i\theta})=1$ for any $m$. Furthermore, from
(\ref{aaa}) we have $A_+=A_-=0$.  So the case 2 in the previous
section applies and $\|\beta,\delta_k\rangle=|k\rangle$.
We finally arrive at the same conclusion as (\ref{limnum}):
$$
    |\beta,\delta_k\rangle \stackrel{\eta\to 1}{\longrightarrow}
    |k\rangle.
$$
Therefore we conclude that the limit to the number states is true
not only for the binomial states ($\mu=1,\
\nu=0$ and $k=M$) but for the more general GBS equations and
all of their eigenstates.

Then we turn to the limit to the coherent and the squeezed states.
As before we let $M\to \infty,\,\eta\to 0$ with fixed
$\eta M =\alpha^2$. However, in the present context we have a whole 
range 
of the parameter $k$, $0\leq k\leq M$, whose limit must be specified,
too. We consider two simple cases.

{\it Case 1}. When $k=K+ p$, where $K=M/2$ for even $M$ and
$(M\pm 1)/2$ for odd $M$, and $p$ is finite. In this case
$(2k-M)$ is a finite integer and $\sqrt{\eta}(2k-M)$ goes
to zero in the limit $\eta\to 0$. 
Multiplying both sides of (\ref{ness}) by $\sqrt{\eta}$
and then taking the limit $M\to \infty,\,\eta\to 0$ with fixed
$\eta M =\alpha^2$ and $n$, we arrive at
\begin{equation}
    \left(\alpha(\mu a+\nu a^{\dagger})-{\alpha^2\over2}\right)
    |\mu,\nu,0,\infty;\delta_k\rangle=0.
    \label{coh1}
\end{equation}
So (\ref{coh1}) becomes
\begin{equation}
   (\mu a +\nu a^{\dagger})|\mu,\nu,\infty;\delta_k\rangle
   ={\alpha\over2}|\mu,\nu,\infty;\delta_k\rangle,
\end{equation}
from which we see the state $|\mu,\nu,0,\infty;\delta_k\rangle$
is a squeezed state. 

{\it Case 2}. When $(2k-M)\propto M$, for example, $k=M-p$ or
$k=p$, where $p$ is finite. In these cases, as $\eta\to0$ and 
$M\to\infty$ for fixed $\eta M=\alpha^2$, $\sqrt{\eta}A_0(2k-M)$
becomes infinite. This conclusion is based on the assumption that
$A_0$ remains finite as $\eta\to0$, which is true provided $\nu\neq0$:
\begin{eqnarray}
  &&  A_0=2\sqrt{1-\eta}\, (\mu+\nu e^{2i\theta})\frac{\Delta}{
      1+|\Delta|^2}+\sqrt{\eta}\cos(2r), \\
  &&  \Delta=-\frac{1}{2\mu}\sqrt{\frac{\eta}{1-\eta}}\pm
      \frac{1}{2\mu}\sqrt{\frac{\eta}{1-\eta}+4\mu\nu}.
\end{eqnarray}
(The exceptional case of $\nu=0$ will be discussed in some detail in 
the next section.)
Therefore in these cases the naive $\eta\to0$ and 
$M\to\infty$ limit does not exist for the GBS. 
In order to define proper limits in these cases we
have to consider the situation in which $\mu$ and $\nu$ are
$\eta$ dependent.

\section{The case $\nu=0$: time evolution of BS}
\setcounter{equation}{0}

In this section we consider the special case $\nu=0$, for which
 no SU(2) rotation is necessary to make $A_-=0$, ie. 
 $D(\zeta)=1$. The eigenvalues and eigenstates
are directly obtained from (\ref{delk}) and (\ref{eista1},
\ref{eista2}) with $A_0=\sqrt{\eta}$ and $A_+=\mu \sqrt{1-\eta}$
\begin{eqnarray}
&&    \delta_k=\frac{1}{2}\sqrt{\eta}(2k-M), \\
&&    |\beta,\delta_k\rangle = D_0 \exp\left\{\mu^{-1}
      \sqrt{\frac{\eta}{1-\eta}}\sqrt{\frac{k-N+1}{M-N+1}}
      J_k^-\right\}|0\rangle \\
&&    \hspace{1.2cm}
      =D_0'\sum_{n=0}^k \left(\begin{array}{c}k\\n\end{array}
      \right)\left(\begin{array}{c}M\\n\end{array}\right)^{
      -1/2}\sqrt{\eta^n(1-\eta)^{k-n}}\mu^{-n}|n\rangle,
      \label{nu0}
\end{eqnarray}
where $D_0$ and $D_0'$ are normalization constants. In
particular, if $\mu=1$, we recover the binomial states
for $k=M$ and find all the other eigenstates of
the operator in (\ref{eig}). 
All these states  for $k\neq M$ are new. If $\mu\equiv
|\mu|e^{i\phi}\neq 1$, we shall see that $|\mu|$ is not essential
but that  its phase is related to  the time evolution of the states
$|\beta,\delta_k\rangle$. In fact $|\mu|$ dependence in 
(\ref{nu0}) can be absorbed by a  new parameter
$\bar{\eta}=\eta/(\eta+|\mu|(1-\eta))$, which also satisfies
$0<\bar{\eta}<1$. So, without loss of generality, we
suppose $|\mu|=1$ in the following. To understand the physical
meaning of the phase of the parameter $\mu=e^{i\phi}$, we consider 
the time evolution of the
states (\ref{nu0}). Suppose that at the initial time $t=0$,
the radiation field is in the state (\ref{nu0}), then at any time $t$,
it is in the state $U(t)|\beta,\delta_k\rangle$, where
$U(t)=e^{-iHt/\hbar}$ is the evolution operator and $H=\omega(N+1/2)$
is the Hamiltonian of the single-mode radiation field. It is obvious 
that
\begin{equation}
     U(t)|\beta,\delta_k\rangle=D_0''\sum_{n=0}^k
     \left(\begin{array}{c}k\\n\end{array}
     \right)\left(\begin{array}{c}M\\n\end{array}\right)^{
     -1/2}\sqrt{\eta^n(1-\eta)^{k-n}}e^{-in(\phi+\omega 
t/\hbar)}|n\rangle,
\end{equation}
from which we see that the phase $\phi$ can be understood as
the shift of the origin of the time.

The state $|\beta,\delta_M\rangle$ is essentially the binomial
state. It can also be understood as the SU(2) coherent states.
In fact, it is not difficult to see that
\begin{equation}
  |\beta,\delta_M\rangle=\exp(\xi' J^+_M - \xi'^* J^-_M)|0\rangle,
  \label{lmdis}
\end{equation}
where $\xi'=-\left[\arctan\sqrt{\eta/(1-\eta)}\right]
e^{i\phi}$. From (\ref{lmdis}) we see that the states
(\ref{nu0}) are the {\it general} SU(2) coherent states due to
the arbitrariness of $\mu$.

It is easy to see that the special cases discussed in the previous 
section (i) $k={M\over2}+p$, (iia) $k=M-p$, (iib) $k=p$, ($p$: finite)
reduce to the coherent states $|\alpha e^{-i\phi}/\sqrt2\rangle$ (i), 
$|\alpha e^{-i\phi}\rangle$ (iia) and to the vacuum $|0\rangle$ 
(iib), respectively, in
the limit $M\to \infty,\,\eta\to 0$ with fixed
$\eta M =\alpha^2$.

\section{Conclusion}
\setcounter{equation}{0}

We have found that, among the three methods for defining the coherent
state of the radiation field, the ladder and displacement
operator methods can be generalized to the study of the BS.
The only exception is the minimum uncertainty method.  This is
understandable since the BS is between the {\it most non-classical}
and the {\it most classical} states, not the most-classical
state (minimum uncertainty state).

On the basis of the analysis of BS we proposed an GBS equation
and solved it exactly. The eigenstates of the GBS equation
corresponding to $M+1$ distinct eigenvalues are all obtained
and their time evolution is discussed. These states range from
the  {\it displaced binomial states} (for $k=M$) to the binomial
states (for $k=0$). They degenerate to the number states and the
coherent and squeezed states in two different limits.
The original BS of Stoler {\it et al} is only an eigenstate
of a special GBS equation ($\nu=0$ and eigenvalue $\sqrt{\eta}M/2$).

Further  investigation of the statistical and phase
properties of the GBS  obtained in this paper will be published
elsewhere.

\bigskip

This work is supported partially by the grant-in-aid for
Scientific Research, Priority Area 231 ``Infinite Analysis'',
Japan Ministry of Education. H.\,C.\,F is grateful to Japan
Society for Promotion of Science (JSPS) for the fellowship.
He is also supported in part by the National Science
Foundation of China.


\end{document}